\documentclass[a4paper,fleqn]{cas-dc}

\usepackage[numbers,sort&compress]{natbib}

\def\tsc#1{\csdef{#1}{\textsc{\lowercase{#1}}\xspace}}
\tsc{WGM}
\tsc{QE}
\tsc{EP}
\tsc{PMS}
\tsc{BEC}
\tsc{DE}

\begin{document}
\let\WriteBookmarks\relax
\def\floatpagepagefraction{1}
\def\textpagefraction{.001}

\shorttitle{Combined optimization ghost imaging based on random speckle field}

\shortauthors{Z. Yang, }

\title [mode = title]{Combined optimization ghost imaging based on random speckle field}


%
\author[1]{Zhiqing Yang}
\author[1]{Cheng Zhou}
\cormark[1]
\ead{zhoucheng91210@163.com}
\author[1]{Gangcheng Wang}
\cormark[1]
\ead{wanggc887@nenu.edu.cn}
\author[2]{Lijun Song}
\cormark[1]
\ead{ccdxslj@126.com}




\affiliation[1]{School of Physics, Northeast Normal University, Changchun 130024, China}
\affiliation[2]{Changchun Institute Of Technology, Changchun 130012, China}
\cortext[cor1]{Corresponding authors}

\begin{abstract}
Ghost imaging is a non local imaging technology, which can obtain target information by measuring the second-order intensity correlation between the reference light field and the target detection light field. However, the current imaging environment requires a large number of measurement data, and the imaging results also have the problems of low image resolution and long reconstruction time. Therefore, using orthogonal methods such as QR decomposition, a variety of optimization methods for speckle patterns are designed combined with Kronecker product,which can help to shorten the imaging time, improve the
imaging quality and image noise resistance.
\end{abstract}




\maketitle

\section{Introduction}

Detection and imaging separation are the imaging features of ghost imaging (GI) which combine optical field coding and computation\cite{shapiro2008computational}. This is a new imaging method that is different from traditional direct detection imaging. Traditional imaging technology has been deeply integrated into various aspects of production and life, and is magnificent. How new imaging technologies showcase their unique features, explore their irreplaceability, and truly replace or compensate for traditional imaging technologies is the focus of research and application promotion personnel. GI, as a new type of imaging technology, has undergone nearly 30 years of development and is also facing the stage from novel demonstration research to engineering application\cite{gibson2020single}. The significance of technology lies in truly solving practical problems for application. Hence, three-dimensional\cite{sun20133d,zhu2014three,gong2016three,sun2016single1}, spectral, polarization, phase, X-ray\cite{han2018review,luo2018orthonormalization}, neutron\cite{he2021single,kingston2020neutron}, THz\cite{chen2020ghost} imaging technology research has been reported in order to promote the application of GI technology.

However, the imaging method that relies on light field encoding and computation has to some extent limited the application of GI, mainly due to the low performance of optical field encoding and the high consumption of computing resources. The lower the performance of optical field encoding, the greater the computational consumption. It can be seen that the optical field coding is very important for the imaging performance and application of GI. Therefore, this paper focuses on the optimization of optical field coding to understand the properties of optical field coding and improve the imaging quality.
At present, the optical field coding used in GI mainly includes random speckle\cite{yang2016scalar}, Hadamard basis\cite{wang2016fast} and Fourier basis matrix. Among them, the hadamard basis and Fourier basis are orthodontic matrices, which can achieve nearly perfect reconstruction of the target object under full sampling, so a lot of optimization work is carried out to improve the quality of under-sampled imaging, especially the Hadamard basis matrix. Since the Hadamard matrix shows worse noise resistance than the random matrix in the experiment, we focus on optimizing the light field encoding of random speckle.


In current research work, the optimization of light field encoding for random speckle mainly focuses on the multi-scale optimization layout of the spatial distribution of the speckle field itself and the orthogonal transformation of the speckle field. Chen et al. explored the possibility of constructing efficient measurement matrices with multiple correlated scale random matrices to improve GI quality\cite{Chen2014}. Wang et al.\cite{Wang2020} improved the hybrid speckle compression computational ghost imaging scheme proposed by Zhou et al.\cite{Zhou2016} to improve the practicality and imaging performance of the scheme. Sun et al. used a multi-scale adaptive light field coding scheme to achieve adaptive high-resolution GI of the part of interest\cite{Sun2016}. Based on the prior statistical information of multi-scale target scenes in Ref. \cite{Chen2014}, Ma et al. further proposed multi-scale push sweep mode ghost imaging lidar via sparsity constraints and pushed it to the engineering application test stage\cite{Ma2019}. Lin et al. have also used multi-scale speckle to carry out related research on anti-scattering media\cite{Lin2023}.

On the other hand, the orthogonalization of speckle fields optimization method mainly aims to optimize the optical field encoding of non orthogonal random speckle into an encoding with orthogonal properties. Ideally, it should maintain its original good noise resistance while possessing the ability to fully recover the orthogonal optical field encoding from full sampling, and further improve the ability to obtain image information under undersampling. zhang et al.\cite{Zhang2014} proposed the processing of optical field coding against orthogonalization, which significantly improved the imaging quality of random speckle, and Gong\cite{Gong2015} conducted experimental verification of this method. Then the two groups carried out the optimization work of quantifying the binomial distribution by random speckle scale, and the noise was further restrained. Zhou et al. designed a mask using the maximum inter-class variance to effectively suppress correlated noise\cite{Zhou2019}. The optimization method of singular value decomposition\cite{Zhang2018,Wang2024}  also shows good performance in light field optimization, and more research work has been carried out in improving image quality and encryption applications. Recently, Gram-Schmidt orthodontic \cite{luo2018orthonormalization} and triangular decomposition optimization\cite{Liu2022} work have been proposed, showing high performance and further research and exploration potential in matrix orthodontic and under-sampling imaging quality improvement.

In order to further clarify the light field coding optimization and thus improve the ghost imaging quality, based on the previous research work above, we proposed a combination optimization ghost imaging scheme based on random speckle fields. This scheme combines Gram-Schmidt orthogonalization and triangular decomposition orthogonalization to optimize the Kronecker product combination of multi-scale and multi-scale speckle patterns, and analyzes the optimized light field. The imaging quality is significantly improved before optimization, which helps to improve the performance of ghost imaging from the root and promote application.

\section{Methods}

Based on the principle of GI, one single pixel detector is selected to obtain the echo signal after the random speckle field $I^{m}(x,y)$ interacts with the target object $O(x,y)$ ($x=1,2,\cdots,r$; $y = 1,2,\cdots,c$). The echo signal $B^{m}~(m=1,2,\cdots,M)$ is collected by the single pixel detector after the modulated random speckle field $I^{m}(x,y)$ irradiates the target $O(x,y)$, where $m$ is the number of measurement times,
\begin{equation}
B^{m}=\iint I^{m}(x,y)O(x,y)dxdy.\label{eq1}
\end{equation}

Then, in the $m-$th ($m=1,2,3,\cdots,M$) measurement, the reconstruction results of the random speckle field obtained through subtracting background second-order correlation can be expressed as:
\begin{eqnarray}
G(x,y) &=& \langle I^{m}(x,y)B^{m} \rangle-\langle I^{m}(x,y) \rangle\langle B^{m} \rangle,\label{eq2}
\end{eqnarray}
where, $\langle  \rangle=\sum_{m=1}^M()$. Due to the non orthogonal nature of the random speckle matrix, the quality of GI based on this is poor. In order to improve the quality, we studied a combination optimization method based on Gram-Schmidt orthogonalization and triangular decomposition in the following work.

\textit{Firstly, Gram-Schmidt orthogonalization scheme can transform the above matrices of speckle patterns for the matrix transformation.} We assume that $M$ measurements will be performed, then $M$ random speckle light field matrices will be converted into column vectors with the same size as the target object, and stored in a matrix $\boldsymbol{X}$ with $N(N=c\times r)$ rows and $M$ columns which $N$ representing the number of pixels of the target object, which can be written as:
\begin{equation}
\boldsymbol{X}=[\varphi_1,\varphi_2, \cdots ,\varphi_M],
\end{equation}
where,
\begin{equation}
\varphi_m=[I^{m}(1,1), \cdots, I^{m}(r,1), \cdots, I^{m}(1,c), \cdots, I^{m}(r,c) ].
\end{equation}

In the matrix $\boldsymbol{X}$, we use $\varphi_m$ to represent each column vector, which is used to store the information about a speckle field and is linearly independent of each other. A set of orthogonal vector groups $\phi_m$ s generated through Gram-Schmidt orthogonalization, in which each column vector represents a optimized speckle light field.
The transformation process can be expressed as:
\begin{align}
\phi_1 &= \varphi_1, \label{eq5}\\
\phi_2 &= \varphi_2-\frac{\phi_1^T\varphi_2}{\phi_1^T\phi_1}\phi_1,\\
\phi_3 &= \varphi_3-\frac{\phi_1^T\varphi_3}{\phi_1^T\phi_1}\phi_1-\frac{\phi_2^T\varphi_3}{\phi_2^T\phi_2}\phi_2,\\
\vdots &= \vdots,\\
\phi_M &= \varphi_M-\frac{\phi_1^T\varphi_M}{\phi_1^T\phi_1}\phi_1-\cdots-\frac{\phi_{M-1}^T\varphi_M}{\phi_{M-1}^T\phi_{M-1}}\phi_{M-1},\label{eq9}
\end{align}
 where, the transformation coefficient can be prescribed by:
\begin{equation}
\xi_{(n,m)}=\frac{\phi_{n}^T\varphi_m}{\phi_{n}^T\phi_{n}},
\end{equation}
where, $n=1,2,3,\cdots,M-1$. Hence, Eq. (\ref{eq9}) can be simplified as:
\begin{equation}
\phi_M = \varphi_M-\xi_{(1,M)}\phi_1-\cdots-\xi_{(M-1,M)}\phi_{M-1},
\end{equation}

For deriving a new column vector, we subtract the projection component of the previous vector on the vector. Obviously, the column vectors in the resulting optimized vector groups ${\phi_m}$by Eqs. (\ref{eq5})-(\ref{eq9}) are pairwise orthogonal. Finally, we achieve normalizated orthogonal basis by introducing the calculation of 2-norm, which can be expressed as:
\begin{equation}\label{ppp}
  \Psi_m=\frac{\phi_m}{\|\phi_m\|}.
\end{equation}

The ultimate vector group is combined in matrix $\boldsymbol{Y}$ and can be expressed as:
\begin{equation}\label{yyy}
  \boldsymbol{Y}=\{\Psi_m\}=[\Psi_1, \Psi_2, \cdots, \Psi_M],~m=1,2,3,\cdots,M.
\end{equation}
where,
\begin{equation}
  \Psi_m=[I_S^m(1,1),\cdots,I_S^m(r,1),\cdots,I_S^m(1,c)  \cdots,I_S^m(r,c)]^{T},
\end{equation}
where, $I_S^m(x,y)$ is the optimized spatial light field of the $m-$th detection  after Gram-Schmidt orthogonal optimization. Then, after substituting it into Eq. (\ref{eq1}), the reconstructed image optimized by Gram-Schmidt orthogonalization can be obtained through the correlation calculation of Eq. (\ref{eq2}).


\textit{Secondly, we use the orthogonal optimization method of triangular decomposition, also referred to as QR decomposition}. The QR decomposition can be expressed as:
\begin{equation}\label{eq15}
\mathbf{X}=\mathbf{Q}\mathbf{R},
\end{equation}
where, $\boldsymbol{Q}$ is the orthogonal matrix, $\boldsymbol{R}$ is the upper triangular matrix. We can use Gram-Schmidt to calculate $\boldsymbol{Q}$.
\begin{equation}\label{yyy}
\boldsymbol{Q}=[\Psi_1, \Psi_2, \cdots, \Psi_M].
\end{equation}
\par So Eq. (\ref{eq15}) can be represented as
\begin{equation}
\boldsymbol{X}=[\varphi_1,\varphi_2, \cdots ,\varphi_M]=[\Psi_1, \Psi_2, \cdots, \Psi_M]\boldsymbol{R}.\label{eq17}
\end{equation}
\par Then, convert Eq. (\ref{eq5})-(\ref{eq9}) to the following form.	
\begin{align}
\varphi_1&=\phi_1={\|\phi_1\|}\Psi_1,\label{eq18}\\
\varphi_2&={\|\phi_2\|}\Psi_2+[\Psi_1,\varphi_2]\Psi_1,\\
\varphi_3&={\|\phi_3\|}\Psi_3+[\Psi_1,\varphi_3]\Psi_1+[\Psi_2,\varphi_3]\Psi_2,\\
\vdots&=\vdots,\nonumber\\
\varphi_M&={\|\phi_M\|}\Psi_M+[\Psi_1,\varphi_M]\Psi_1+[\Psi_2,\varphi_M]\Psi_2+\nonumber\\
&+[\Psi_{M-1},\varphi_M]\Psi_{M-1}.\label{eq22}
\end{align}

\par Express Eq. (\ref{eq18})-(\ref{eq22}) in matrix form according to Eq. (\ref{eq17}).
\begin{align}
\boldsymbol{X}&=[\varphi_1,\varphi_2, \cdots ,\varphi_M]=[\Psi_1, \Psi_2, \cdots, \Psi_M]\nonumber\\
&\begin{bmatrix}
\|\phi_1\|&[\Psi_1,\varphi_2]&\dots&[\Psi_1,\varphi_M]\\
&\|\phi_2\|&\dots&[\Psi_2,\varphi_M]\\
&&\ddots&\vdots\\
&&&\|\phi_M\|
\end{bmatrix}.
\end{align}
\par So we get the upper triangular matrix $R$ is
\begin{align}
\boldsymbol{R}&=\{\eta_m\}=[\eta_1,\eta_2,\cdots,\eta_M]^{T}\nonumber\\&=
\begin{bmatrix}
\|\phi_1\|&[\Psi_1,\varphi_2]&\dots&[\Psi_1,\varphi_M]\\
&\|\phi_2\|&\dots&[\Psi_2,\varphi_M]\\
&&\ddots&\vdots\\
&&&\|\phi_M\|
\end{bmatrix},
\end{align}
where, $m$ = 1,2,3,$\cdots$,$M$ and
\begin{equation}
\eta_m=
[I_R^m(1,1),\cdots ,I_R^m(r,1),\cdots,I_R^m(1,c),\cdots ,I_R^m(r,c)]^{T},
\end{equation}
where, $I_R^m(x,y)$ is the optimized spatial light field of the $m-$th detection by QR decomposition. It should be noted that in QR decomposition, the matrix $\boldsymbol{X}$ has a unique solution for $\boldsymbol{Q}$ and $\boldsymbol{R}$.


\textit{Third, we perform joint optimization of random speckle through Kronecker product, Gram-Schmidt orthogonalization, and QR decomposition.} Because Gram-Schmidt orthogonal and QR decomposition only show high quality image under full sampling, the image quality is poor under under-sampling. Therefore, we use QR decomposition and Gram-Schmidt orthogonalization to optimize random speckle field, aiming to improve the image quality under the condition of under-sampling. Hence, we specifically designed several optimization methods, as follows:

(1) The $\boldsymbol{R}$ matrix is obtained by QR decomposition of speckle matrix $\boldsymbol{X}$, and then the $\boldsymbol{R}$ matrix is orthogonalized by Gram-Schmidt, abbreviated as $\textrm{RS}_{GI}$.

(2) Generate a small size matrix $\boldsymbol{X_1}$ of size ($N, \frac{N}{2}$), and expand it to a large matrix $\boldsymbol{X}$ of size ($N, N$) by Kronick product, and then perform the operation of method (1), referred to as $\textrm{K1RS}_{GI}$.

(3)Generate two small-sized random matrices $\boldsymbol{X_1}$ and $\boldsymbol{X_2}$ with dimensions ($\frac{N}{8}, \frac{N}{8}$) and ($\frac{N}{2}, (\frac{N}{2}-\frac{N}{8})$), and use Kronecker product to extrapolate matrix $\boldsymbol{X_2}$ to matrix $\boldsymbol{X_3}$ with dimensions ($\frac{N}{8}, \frac{N}{2}$), forming a matrix $\boldsymbol{X_4}$ with dimensions ($\frac{N}{2}, \frac{N}{2}$) from $\boldsymbol{X_2}$ and $\boldsymbol{X_3}$. Then, use Kronecker product to expand to a large matrix $\boldsymbol{X}$ with dimensions ($N, N$).  Finally, perform the operation of method (1), referred to as $\textrm{K2RS}_{GI}$.

(4)Generate two small-sized random matrices $\boldsymbol{X_1}$ and $\boldsymbol{X_2}$ with dimensions ($\frac{N}{4}, \frac{N}{4}$) and ($\frac{N}{2}, (\frac{N}{2}-\frac{N}{4})$), and use Kronecker product to extrapolate matrix $\boldsymbol{X_2}$ to matrix $\boldsymbol{X_3}$ with dimensions ($\frac{N}{4}, \frac{N}{2}$), forming a matrix $\boldsymbol{X_4}$ with dimensions ($\frac{N}{2}, \frac{N}{2}$) from $\boldsymbol{X_2}$ and $\boldsymbol{X_3}$. Then, use Kronecker product to expand to a large matrix $\boldsymbol{X}$ with dimensions ($N, N$).  Finally, perform the operation of method (1), referred to as $\textrm{K3RS}_{GI}$.

\section{Results}
In order to verify the imaging performance of the proposed speckle field optimization scheme, we carried out numerical simulation studies by using QR-decomposed $\boldsymbol{Q}$ matrix ($\textrm{Q}_{GI}$), R matrix ($\textrm{R}_{GI}$), Gram-Schmidt orthographic matrix ($\textrm{S}_{GI}$), $\textrm{RS}_{GI}$, $\textrm{K1RS}_{GI}$, $\textrm{K2RS}_{GI}$ and $\textrm{K3RS}_{GI}$ methods respectively.

\begin{figure}[htbp]
\centering
\includegraphics[width=8cm]{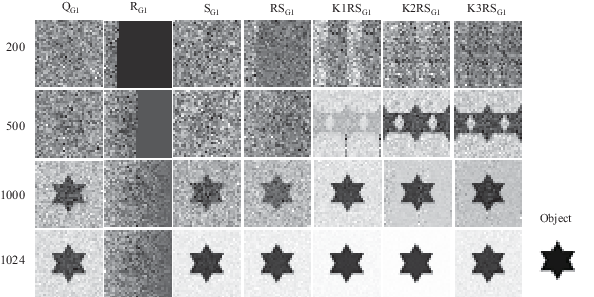}
\caption{Numerical simulation results of $\textrm{Q}_{GI}$, $\textrm{R}_{GI}$, $\textrm{S}_{GI}$, $\textrm{RS}_{GI}$, $\textrm{K1RS}_{GI}$, $\textrm{K2RS}_{GI}$ and $\textrm{K3RS}_{GI}$ schemes with different measurement times.}
\label{star}
\end{figure}

First, we perform numerical simulation for sparse target, the sparse object size is $32 \textrm{pixel}\times 32 \textrm{pixel}$  black hexagonal on white background, as shown in Fig.~\ref{star} Object. And, the numerical simulation results of $\textrm{Q}_{GI}$, $\textrm{R}_{GI}$, $\textrm{S}_{GI}$, $\textrm{RS}_{GI}$, $\textrm{K1RS}_{GI}$, $\textrm{K2RS}_{GI}$ and $\textrm{K3RS}_{GI}$ schemes with different measurement times are shown in Fig.~\ref{star} . It can be seen from the results that the orthogonal $\boldsymbol{Q}$ matrix obtained by QR decomposition can obtain a better reconstructed image at 1024 times of full sampling (the first column in Fig.~\ref{star}), while the upper triangle $\boldsymbol{R}$ matrix obtained by QR decomposition has poor quality because it is not orthographic and the lower triangle is 0. This case, some areas will have the same value and no information can be obtained in the low sampling (the second column in Fig.~\ref{star}). Since the field optimization performance of Gram-Schmidt orthographic is similar to that of QR-decomposed orthogonal matrix $\boldsymbol{Q}$, a similar high quality reconstructed image can be obtained in full sampling, as shown in the third column in Fig.~\ref{star}.

Then, we verify the four optimization methods proposed in this paper. First, Gram-Schmidt orthonoralization of QR decomposition $\boldsymbol{R}$ matrix is performed to obtain the results in the fourth column of Fig.~\ref{star}. From the results, it can be seen that the $\textrm{RS}_{GI}$ optimized by $\boldsymbol{R}$ matrix achieves the reconstruction results of high quality, reaching the imaging level of $\textrm{Q}_{GI}$ and $\textrm{S}_{GI}$. Moreover, the image quality of the three schemes $\textrm{K1RS}_{GI}$, $\textrm{K2RS}_{GI}$ and $\textrm{K3RS}_{GI}$ expanded by Kronick product has been greatly improved, especially under the condition of under-sampling. However, although hexagonal can be clearly seen at 500 times, there is aliasing, which gradually disappears after the number of times increases. Visually, $\textrm{K2RS}_{GI}$ works best.

\begin{figure}[htbp]
\centering
\includegraphics[width=8cm]{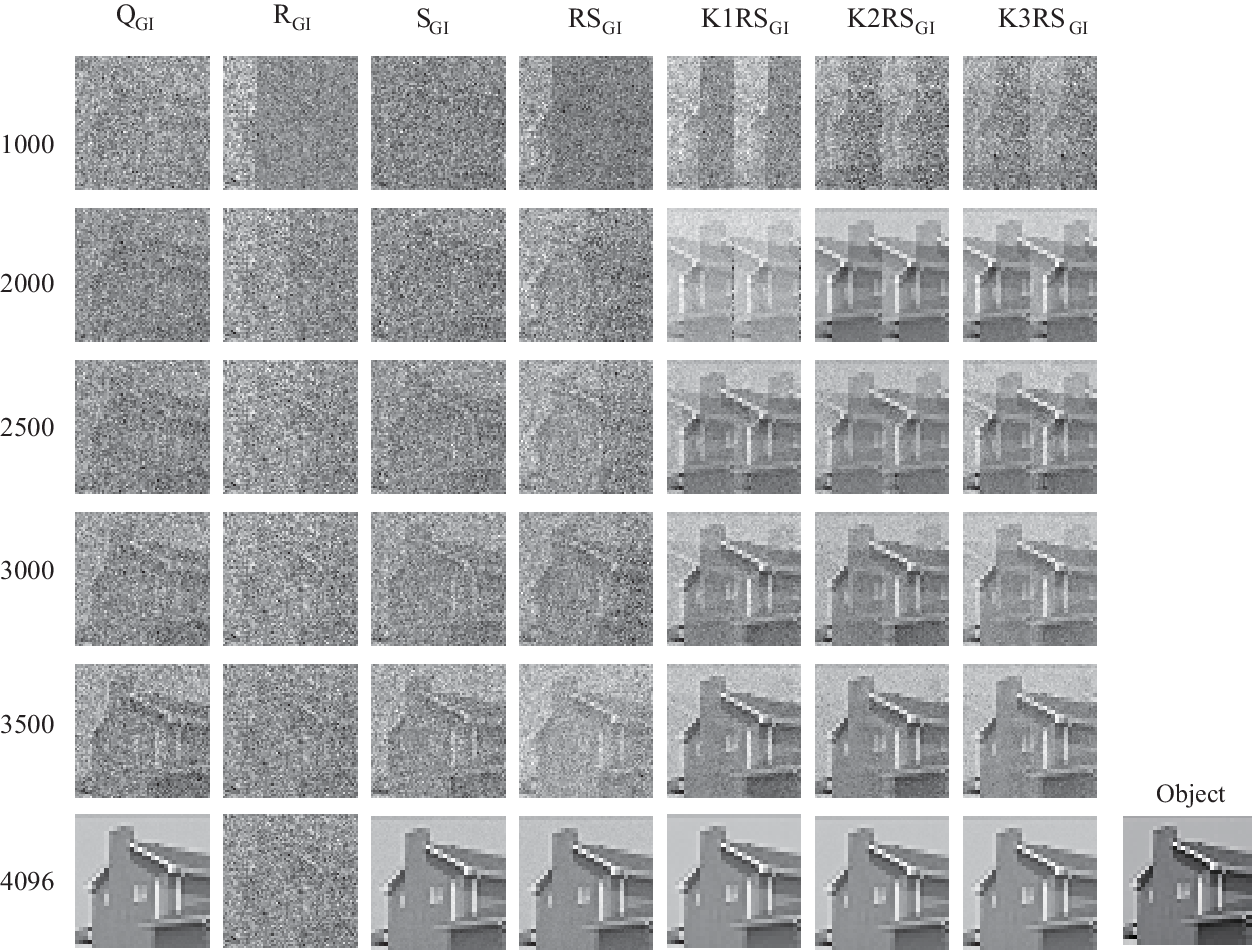}
\caption{Numerical simulation results of $\textrm{Q}_{GI}$, $\textrm{R}_{GI}$, $\textrm{S}_{GI}$, $\textrm{RS}_{GI}$,  $\textrm{K1RS}_{GI}$, $\textrm{K2RS}_{GI}$ and $\textrm{K3RS}_{GI}$ schemes with different measurement times.}
\label{house}
\end{figure}

In order to further prove the effectiveness of grayscale object and provide a solution to solve the problem that the long measurement time of CGI cannot meet the application requirements of fast imaging, we selected a $64\textrm{pixel}\times 64\textrm{pixel}$ house as the object of simulation, and carried out the simulation experiment of the 7-clock method respectively. The image comparison of the 7-clock method is the same as that of the hexagonal case. The three schemes $\textrm{K1RS}_{GI}$, $\textrm{K2RS}_{GI}$ and $\textrm{K3RS}_{GI}$ optimized by Kronick product are still the ones with the greatest improvement in image quality. The house features can be obtained at a lower number of measurements and aliasing still exists, but the reconstructed images with higher quality and no information aliasing can be obtained when the number of measurements is 3000, as shown in Fig.~\ref{house}.

In order to numerically compare the results of the 7-clock method more accurately, we introduce the Mean Square Error (MSE) and the Peak Signal to Noise ratio (PSNR) as measures of image quality. MSE is used to reflect the difference between the result $G(x,y)$ and the object $O(x,y)$, so as to determine the distortion degree of the reconstructed images, expressed as
\begin{eqnarray}
\mathrm{MSE}=\frac{1}N\sum \limits _{x,y} ^{N} [G(x,y)-O(x,y)]^{2}
\end{eqnarray}
where $N$ represents the number of pixels of $G(x,y)$ and $O(x,y)$.
 \par
\begin{figure}[htbp]
 	\centering
 	\includegraphics[width=4cm]{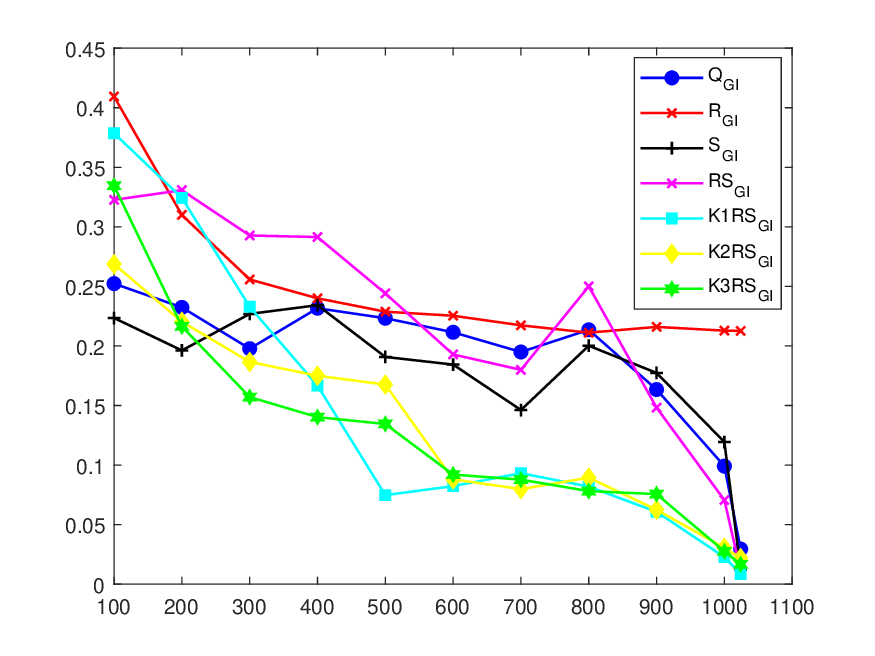}
 	\includegraphics[width=4cm]{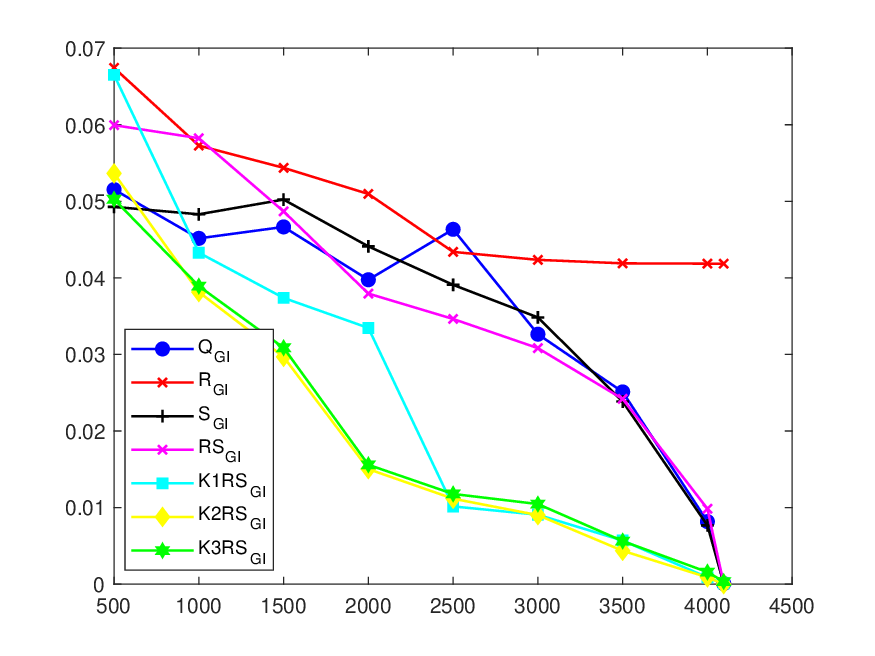}
 	\caption{The MSEs of $\textrm{Q}_{GI}$, $\textrm{R}_{GI}$, $\textrm{S}_{GI}$, $\textrm{RS}_{GI}$, $\textrm{K1RS}_{GI}$, $\textrm{K2RS}_{GI}$ and $\textrm{K3RS}_{GI}$ method with different measurement times. (a) The simulation results of the hexagon. (b) The simulation results of the house}\label{mse}
\end{figure}

 When evaluating image quality, the smaller the MSE value, the better the image quality to be evaluated, and the higher the similarity between the reconstructed results and the target object. The MSEs curve of the $\textrm{Q}_{GI}$, $\textrm{R}_{GI}$, $\textrm{S}_{GI}$, $\textrm{RS}_{GI}$, $\textrm{K1RS}_{GI}$, $\textrm{K2RS}_{GI}$ and $\textrm{K3RS}_{GI}$ method are shown in Fig.~\ref{mse},  where Fig.~\ref{mse}(a) is the simulation result of the image hexagon image with 1024 measurements, and  Fig.~\ref{mse}(b) is the result of the house image with 4096 measurements. It is obvious that the MSEs of all schemes generally decrease with the increase of the number of measurements and the performance of $\textrm{K1RS}_{GI}$, $\textrm{K2RS}_{GI}$ and $\textrm{K3RS}_{GI}$ schemes are better than others, such as in the case of full sampling, the MSEs reaches the lowest value and almost achieves the perfect restoration of the image. Overall, the MSEs of $\textrm{Q}_{GI}$, $\textrm{R}_{GI}$, $\textrm{S}_{GI}$, $\textrm{RS}_{GI}$ schemes are not as low as those of $\textrm{K1RS}_{GI}$, $\textrm{K2RS}_{GI}$ and $\textrm{K3RS}_{GI}$ method for the same measurement number, and the numerical performance of the $\textrm{Q}_{GI}$ scheme is the worst, where even in the case of too many samples, the MSE value are still too high. Meanwhile, by comparing Fig.~\ref{mse}(a) and Fig.~\ref{mse}(b), we found that the larger the pixels of the target, the better the reconstruction results of $\textrm{K1RS}_{GI}$, $\textrm{K2RS}_{GI}$ and $\textrm{K3RS}_{GI}$, indicating that it is suitable for the measurement and restoration of large images to achieve high-quality imaging in a short time.
\begin{figure}[htbp]
	\centering
	\includegraphics[width=4cm]{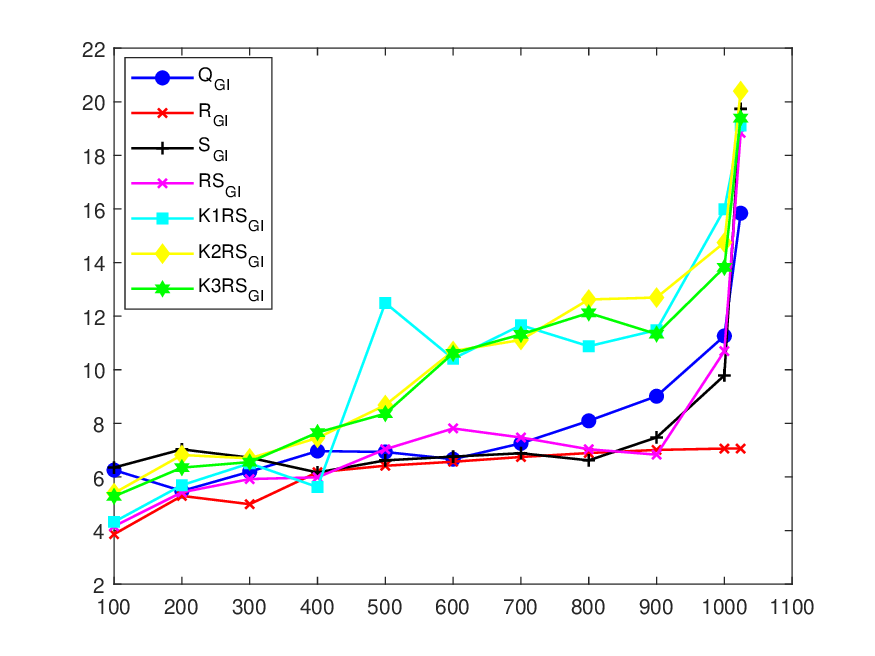}
	\includegraphics[width=4cm]{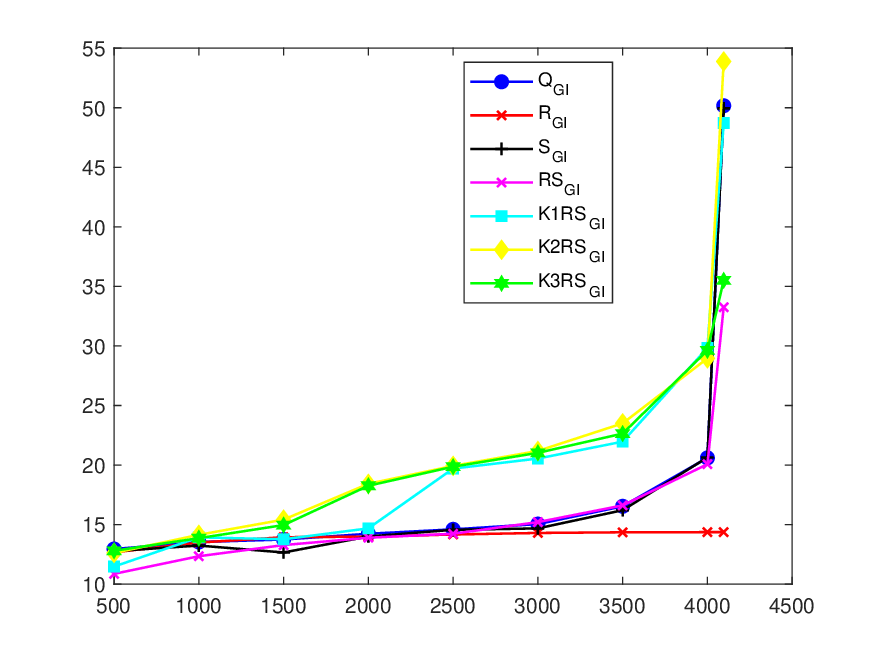}
	\caption{The PSNRs of $\textrm{Q}_{GI}$, $\textrm{R}_{GI}$, $\textrm{S}_{GI}$, $\textrm{RS}_{GI}$, $\textrm{K1RS}_{GI}$, $\textrm{K2RS}_{GI}$ and $\textrm{K3RS}_{GI}$ method with different measurement times. (a) The simulation results of the hexagon. (b) The simulation results of the house}\label{psnr}
\end{figure}
\par As the most common and widely used objective standard for evaluating images, PNSR is used to measure whether reconstruction can get good results. The calculation of PSNR depends on MSE, usually expressed in logarithmic form, which can be represented as
 \begin{eqnarray}
 	\mathrm{PSNR}=10\times\log_{10}[{\frac{(2^{u}-1)^2}{\mathrm{MSE}}}]
 \end{eqnarray}
where $u$ represents the gray scale of pixels. Generally for binary images, $u$=1.
\par Therefore, the larger the MSE, the lower the PSNR, and the greater the distortion of the result image, the lower the similarity with the target. Generally speaking, when the PSNR is between 20dB and 30dB, the human eye can detect the difference between the processed image and the original image; when PSNR is greater than 30dB, it is difficult for human eyes to detect the difference between images; when PSNR is close to 50dB, it means that the processed image has only a small error. Fig.~\ref{psnr} lists the PSNRs curves of 7-clock method in numerical simulations. On the whole, it is obvious that the PNSRs values of all schemes increase slowly with the increases of simulation times in the case of low sampling, where the gap is not obvious. However, as the simulation is more than half completed, the PSNRs values of $\textrm{K1RS}_{GI}$, $\textrm{K2RS}_{GI}$ and $\textrm{K3RS}_{GI}$ show a greater rate of increase with $\textrm{K2RS}_{GI}$ being the best. Even when simulating the house image, the PSNR of $\textrm{K2RS}_{GI}$ reaches nearly 55dB at near full sampling which realises the high restoration of the image. Meanwhile, among $\textrm{Q}_{GI}$, $\textrm{R}_{GI}$, $\textrm{S}_{GI}$ and $\textrm{RS}_{GI}$, $\textrm{Q}_{GI}$ shows higher performance, and is the same as MSEs curves, while $\textrm{R}_{GI}$ scheme is poor.
\begin{figure}[htbp]
	\centering
	\includegraphics[width=4cm]{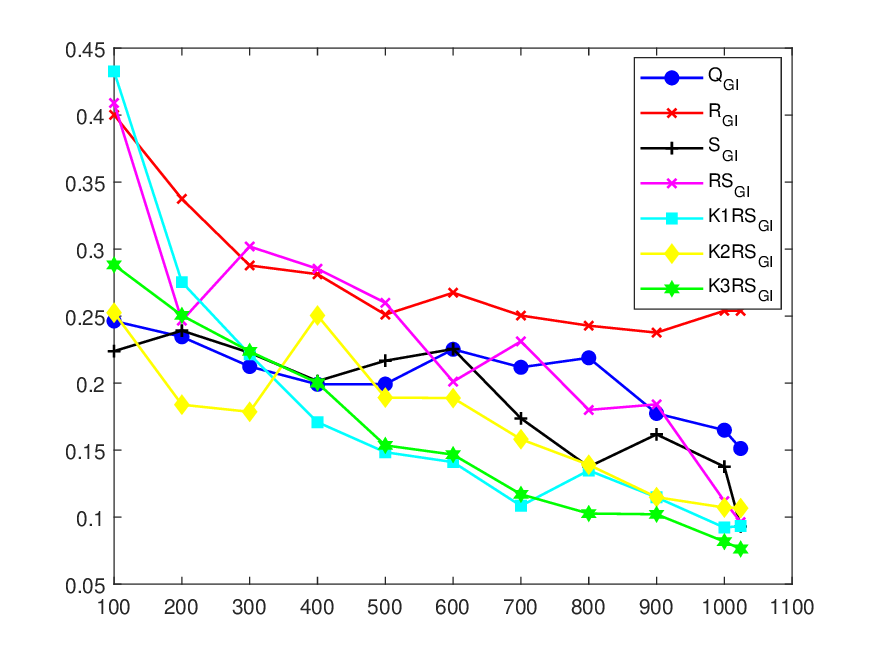}
	\includegraphics[width=4cm]{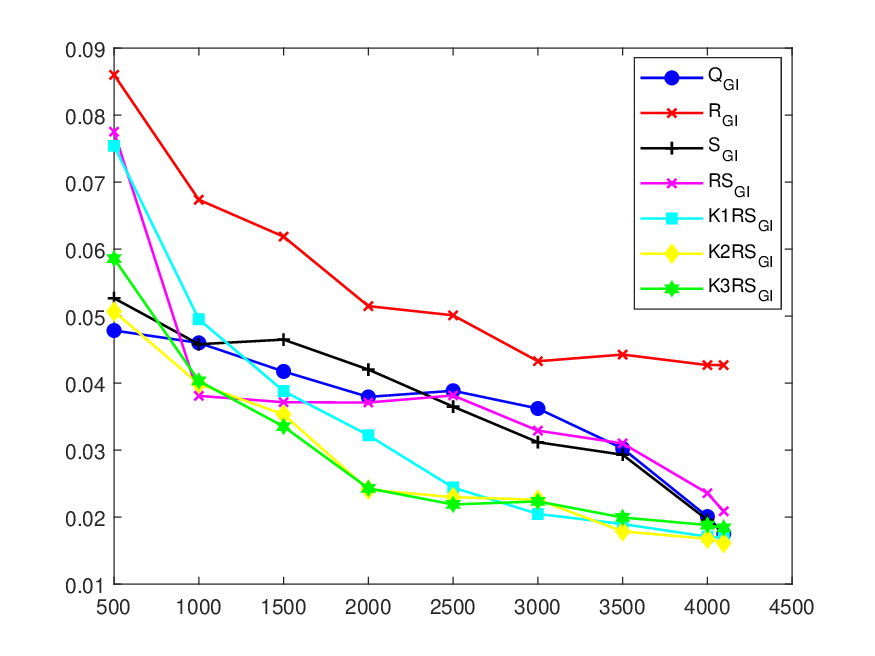}
	\caption{The MSEs of $\textrm{Q}_{GI}$, $\textrm{R}_{GI}$, $\textrm{S}_{GI}$, $\textrm{RS}_{GI}$, $\textrm{K1RS}_{GI}$, $\textrm{K2RS}_{GI}$ and $\textrm{K3RS}_{GI}$ method with the same Gaussian white noise under different measurement times. (a) The simulation results of the hexagon. (b) The simulation results of the house}\label{noise}
\end{figure}

In addition, by setting ratio of signal power to noise power to 15dB, we added the same Gaussian white noise to the "hexagon" and "house" respectively. The MSE value curve of the generated results is shown in the Fig. ~\ref{noise}. According to the results we can get that with the signal has the same slight loss and interference, the performance of $\textrm{K1RS}_{GI}$, $\textrm{K2RS}_{GI}$ and $\textrm{K3RS}_{GI}$
schemes are still better than the other four methods, with higher imaging quality, at the same time the noise immunity of $\textrm{K3RS}_{GI}$ is best.

\section{Conclusion}
 This paper proposes a combined optimization scheme based on Spatial light field transformation. In this scheme, Schmidt orthogonalization (unity), orthogonal trigonometric decomposition and Kronecker product method are used to transform the spatial light field matrix. The optimized spatial light field has the advantages of low complexity and low computational difficulty. We use several optimized light field matrices instead of the original random light field to interact with the object respectively. Theoretical derivation and numerical simulation results verify the effectiveness of the scheme. In the evaluation and analysis of the calculation results, the method of combining subjective evaluation and objective evaluation is adopted. The results show that, compared with the spatial light field without combination optimization, this scheme reduces the amount of redundant calculation in the imaging process, achieves the purpose of obtaining high-quality reconstruction results under low sampling to a certain extent, improves the imaging efficiency, and improves the flexibility of correlation imaging, which provides new thinking for its application in the real field, and has certain practical significance.

\section*{Acknowledgments}

This work is supported by the Science and Technology Development Plan Project of Jilin Province, China (Grant no. 20220204134YY); the National Natural Science Foundation of China (Grant no. 62301140); the Project of the Education Department of Jilin Province (Grant no. JJKH202312 92KJ) and Program for Science and Technology Development of Changchun City (Grant no. 23YQ11); the Innovation and Entrepreneurship Talent Funding project of Jilin Province (Grant No. 2023RY17) and the project of Jilin Provincial Development and Reform Commission (Grant No. 2023C042-4)
\section*{Disclosures}
The authors declare that there are no conflicts of interest related to this paper.


\bibliographystyle{elsarticle-num}


\begin{thebibliography}{10}
\expandafter\ifx\csname url\endcsname\relax
  \def\url#1{\texttt{#1}}\fi
\expandafter\ifx\csname urlprefix\endcsname\relax\def\urlprefix{URL }\fi
\expandafter\ifx\csname href\endcsname\relax
  \def\href#1#2{#2} \def\path#1{#1}\fi

\bibitem{shapiro2008computational}
J.~H. Shapiro, Computational ghost imaging, Physical Review A 78~(6) (2008)
  061802.

\bibitem{gibson2020single}
G.~M. Gibson, S.~D. Johnson, M.~J. Padgett, Single-pixel imaging 12 years on: a
  review, Optics Express 28~(19) (2020) 28190--28208.

\bibitem{sun20133d}
B.~Sun, M.~P. Edgar, R.~Bowman, L.~E. Vittert, S.~Welsh, A.~Bowman, M.~J.
  Padgett, 3d computational imaging with single-pixel detectors, Science
  340~(6134) (2013) 844--847.

\bibitem{zhu2014three}
Y.~Zhu, J.~Shi, H.~Li, G.~Zeng, Three-dimensional ghost imaging based on
  periodic diffraction correlation imaging, Chinese Optics Letters 12~(7)
  (2014) 071101.

\bibitem{gong2016three}
W.~Gong, C.~Zhao, H.~Yu, M.~Chen, W.~Xu, S.~Han, Three-dimensional ghost
  imaging lidar via sparsity constraint, Scientific reports 6~(1) (2016) 1--6.

\bibitem{sun2016single1}
M.-J. Sun, M.~P. Edgar, G.~M. Gibson, B.~Sun, N.~Radwell, R.~Lamb, M.~J.
  Padgett, Single-pixel three-dimensional imaging with time-based depth
  resolution, Nature communications 7~(1) (2016) 1--6.

\bibitem{han2018review}
S.~Han, H.~Yu, X.~Shen, H.~Liu, W.~Gong, Z.~Liu, A review of ghost imaging via
  sparsity constraints, Applied Sciences 8~(8) (2018) 1379.

\bibitem{luo2018orthonormalization}
B.~Luo, P.~Yin, L.~Yin, G.~Wu, H.~Guo, Orthonormalization method in ghost
  imaging, Optics express 26~(18) (2018) 23093--23106.

\bibitem{he2021single}
Y.-H. He, Y.-Y. Huang, Z.-R. Zeng, Y.-F. Li, J.-H. Tan, L.-M. Chen, L.-A. Wu,
  M.-F. Li, B.-G. Quan, S.-L. Wang, T.-J. Liang, Single-pixel imaging with
  neutrons, Science Bulletin 66~(2) (2021) 133--138.

\bibitem{kingston2020neutron}
A.~M. Kingston, G.~R. Myers, D.~Pelliccia, F.~Salvemini, J.~J. Bevitt,
  U.~Garbe, D.~M. Paganin, Neutron ghost imaging, Physical Review A 101~(5)
  (2020) 053844.

\bibitem{chen2020ghost}
S.-C. Chen, Z.~Feng, J.~Li, W.~Tan, L.-H. Du, J.~Cai, Y.~Ma, K.~He, H.~Ding,
  Z.-H. Zhai, Z.-R. Li, C.-W. Qiu, X.-C. Zhang, L.-G. Zhu, Ghost spintronic
  thz-emitter-array microscope, Light: Science \& Applications 9~(1) (2020)
  1--9.

\bibitem{yang2016scalar}
C.~Yang, C.~Wang, J.~Guan, C.~Zhang, S.~Guo, W.~Gong, F.~Gao,
  Scalar-matrix-structured ghost imaging, Photonics Research 4~(6) (2016)
  281--285.

\bibitem{wang2016fast}
L.~Wang, S.~Zhao, Fast reconstructed and high-quality ghost imaging with fast
  walsh--hadamard transform, Photonics Research 4~(6) (2016) 240--244.

\bibitem{Chen2014}
M.~Chen, E.~Li, S.~Han, Application of multi-correlation-scale measurement
  matrices in ghost imaging via sparsity constraints, Applied optics 53~(13)
  (2014) 2924--2928.

\bibitem{Wang2020}
X.~Wang, Y.~Tao, F.~Yang, Y.~Zhang, An effective compressive computational
  ghost imaging with hybrid speckle pattern, Optics Communications 454 (2020)
  124470.

\bibitem{Zhou2016}
C.~Zhou, H.~Huang, B.~Liu, L.~Song, Hybrid speckle-pattern compressive
  computational ghost imaging, Acta Optica Sinica 36~(9) (2016) 0911001.

\bibitem{Sun2016}
S.~Sun, W.-T. Liu, H.-Z. Lin, E.-F. Zhang, J.-Y. Liu, Q.~Li, P.-X. Chen,
  Multi-scale adaptive computational ghost imaging, Scientific reports 6~(1)
  (2016) 37013.

\bibitem{Ma2019}
S.~Ma, C.~Hu, C.~Wang, Z.~Liu, S.~Han, Multi-scale ghost imaging lidar via
  sparsity constraints using push-broom scanning, Optics Communications 448
  (2019) 89--92.

\bibitem{Lin2023}
L.-X. Lin, J.~Cao, D.~Zhou, Q.~Hao, Scattering medium-robust computational
  ghost imaging with random superimposed-speckle patterns, Optics
  Communications 529 (2023) 129083.

\bibitem{Zhang2014}
C.~Zhang, S.~Guo, J.~Cao, J.~Guan, F.~Gao, Object reconstitution using
  pseudo-inverse for ghost imaging, Optics Express 22~(24) (2014) 30063--30073.

\bibitem{Gong2015}
W.~Gong, High-resolution pseudo-inverse ghost imaging, Photonics Research 3~(5)
  (2015) 234--237.

\bibitem{Zhou2019}
Y.~Zhou, S.-X. Guo, F.~Zhong, T.~Zhang, Mask-based denoising scheme for ghost
  imaging, Chinese Physics B 28~(8) (2019) 084204.

\bibitem{Zhang2018}
X.~Zhang, X.~Meng, X.~Yang, Y.~Wang, Y.~Yin, X.~Li, X.~Peng, W.~He, G.~Dong,
  H.~Chen, Singular value decomposition ghost imaging, Optics Express 26~(10)
  (2018) 12948--12958.

\bibitem{Wang2024}
H.~Wang, X.-Q. Wang, C.~Gao, X.~Liu, Y.~Wang, H.~Zhao, Z.-H. Yao, High-quality
  computational ghost imaging with multi-scale light fields optimization,
  Optics \& Laser Technology 170 (2024) 110196.

\bibitem{Liu2022}
J.-F. Liu, L.~Wang, S.-M. Zhao, Orthogonal-triangular decomposition ghost
  imaging, Chinese Physics B 31~(8) (2022) 084202.

\end{thebibliography}

\end{document}